\documentstyle[11pt,paspconf,psfig]{article}

\begin{document}
%----------------------------------------------------------------------
\title{Mixing and Transfer of Elements by Interactions and
Mergers}
\author{Fran\c{c}oise Combes}
\affil{Observatoire de Paris, 61 Av. de l'Observatoire, 
F-75014 Paris, France} 
%----------------------------------------------------------------------
\begin{abstract}
Galaxy interactions produce strong torques that generate 
radial gas flows. There are two opposite tendencies of these
flows, regarding abundance gradients: homogeneization and
gradient flattening, and enhanced star formation in the center,
and gradient steepening. Mechanisms to create abundance
gradients are discussed, and comparison with observations
is detailed, concerning spirals (such as collisional rings,
starbursts..) or ellipticals (formed or not in mergers).
A review of N-body simulations predictions, taken into
account star formation, is also presented.
\end{abstract}

%----------------------------------------------------------------------
% KEYWORDS SHOULD BE INCLUDED, BUT THEY ARE NOT PRINTED IN THE HARDCOPY!

%----------------------------------------------------------------------
\section{Introduction}
 In the mixing and transfer of elements, gas and stars have completely
different roles. The stellar component is not dissipative, and its
density in phase space cannot increase, so that dynamical events
like interactions or mergers tend to homogeneize the system towards
a well-mixed state. On the contrary, gas dissipates energy through
radiation, can concentrate with large proportions towards the nuclei,
and trigger starbursts; since the star formation rate is not linear
with gas density, this tends to re-build abundance gradients.
Therefore, we have to consider these two opposite behaviours
in the dynamics of interactions of galaxies and mergers.

%----------------------------------------------------------------------
\section{Mechanisms to create abundance gradients}
%----------------------------------

Exponential abundance gradients are generally observed
in spiral galaxy disks. The explanation is strongly linked
to the building of an exponential radial distribution
of the density. If the latter is obtained through 
dynamical processes, then the abundance gradient will follow
from the fact that the star formation is a non-linear
function of gas density (e.g. the Schmidt law, where
the SF rate is $\propto \rho_g^n$, with $n>1$).

\subsection{Theory of viscous disks}

It is now well known in the theory of viscous accretion disks
that the main effects of viscosity is to move angular-momentum
outward and mass inward. In the particular hypothesis of
equal time-scales between viscosity and star formation,
i.e. $\tau_{\nu} \sim \tau_*$, it can be shown that the
final surface density of the stellar disk is exponential
(Lin \& Pringle 1987a). This result is independent of
the shape of the rotation curve, and can be obtained as
well in the two extreme cases of a keplerian or flat
rotation curves. 
In this context, Tsujimoto et al (1995) determined
that an exponential distribution of metallicity
is also obtained.

%----------------------------------
\subsection{Gravity torques}

What is the nature of the viscosity, effective in galactic
disks? Normal viscosity is not efficient, due to the
very low density of the gas. One could think of
macroturbulent viscosity, but the time-scales are
longer than the Hubble time at large radii, and could 
be effective only inside the central 1kpc. Instead,
if the galaxy disks develop non-axisymmetric density waves
such as spirals or bars, gravity torques are then very
effective at transfering the angular momentum outward.
 This led Lin \& Pringle (19987b) to propose a 
prescription for an effective kinematic viscosity
for self-gravitating disk undergoing gravitational
instabilities.

%----------------------------------
\begin{figure}[t]
% psfile=#1 vsize=#2 angle=#3 hscale=#4 vscale=#5 hoffset=#6 voffset=#7
%\plotfiddle{combes_qc_f1.ps}{65truemm}{-90}{50}{50}{0}{0}
\psfig{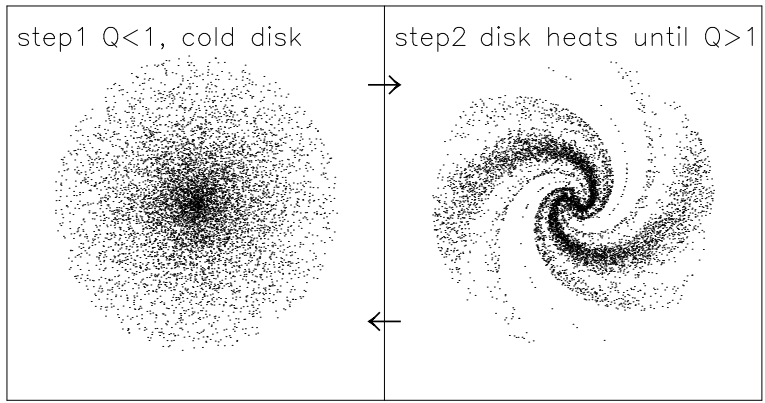}
\caption{Gravitational instabilities regulate
the transfer of angular momentum, through a feedback process,
so that we can talk of "gravitational viscosity": in the first step,
the disk is cold ($Q < 1$), and therefore unstable to spiral waves; second,
the disk has developped waves, that create non-axisymmetry and
gravity torques, that transfer the angular momentum outwards
(trailing waves). The waves heat the disk, until $Q \sim 1$. Then
a disk with only a stellar component will remain stable, while
a gaseous disk can cool back to step 1.}
\end{figure}

The basis of this prescription is described in figure 1.
Gravitational instabilities are suppressed at small scales
through the local velocity dispersion $c$, and at large
scale by rotation. The corresponding limiting scales
are the Jeans scale for a 2D disk $\lambda_J \sim c^2/(G \mu)$,
and $\lambda_c \sim G \mu/\kappa^2$, where $\mu$ is the 
disk surface density, and $\kappa$ the epicyclic frequency.
Scales between $\lambda_J$ and $\lambda_c$ are unstable,
unless $c$ is larger than $\pi G \mu/\kappa$, or the Toomre
$Q = { {c \kappa} \over {\pi G \mu} }$ is larger than 1.
 If the disk is cold at the beginning (general case for the gas),
instabilities set in, which heat the disk until $Q \sim 1$,
and those instabilities provide the necessary angular momentum
tranfer, or viscosity, to concentrate the mass. Since
the size of the region over which angular momentum is tranferred
is $\sim \lambda_c$, and the time-scale is a rotation period,
the effective kinematic viscosity is $\nu \sim \lambda_c^2  \Omega$,
and the typical viscous time 
$\tau_{\nu} \sim R^2 \Omega^3 / (G^2 \mu^2)$.

Why should there be approximate agreement between the 
two time-scales, viscosity and star-formation?
This comes from the fact that the two processes depend
exactly on the same physical mechanism, i.e. gravitational
instabilities. As shown empirically by Kennicutt (1989),
the Toomre parameter $Q$ appears to control star-formation
in spiral disks. Therefore, if the regulating instabilities
have time to develop, one can expect that 
 $\tau_{\nu} \sim \tau_*$, as required for exponential
light and metallicity distribution.

This balance could however be broken in two cases:

\begin{itemize}

\item either the gas flows are too rapid with respect
to star-formation; this can occur in the case of
a strong bar instability just settling. We have then
 $\tau_{\nu} <<  \tau_*$, and the abundance gradients
are smoothed out (Friedli \& Benz 1995, Friedli et al 1994,
Martin \& Roy 1994, 1995).

\item or a starburst is triggered, which makes $\tau_*$
small again, and the gradients are enhanced; the competition between
enhanced star formation and gas flows has been studied by
Edmunds \& Greenhow (1995).

\end{itemize}

%----------------------------------
% TO ENCAPSULATE TWO FIGURES SIDE BY SIDE
%\begin{figure}
%\plottwo{lastname_fig2a.ps}{lastname_fig2b.ps}
%\caption{Caption of Figs.~2a and 2b}
%\end{figure}

%----------------------------------------------------------------------
\section{Observations}

\subsection{Interacting galaxies}

There is a category of interacting galaxies where the
dynamics and star formation processes should be very
simple, and therefore easy to interprete and quantify,
they are the collisional ring galaxies, which prototype
is the Cartwheel (e.g. Appleton \& Struck-Marcell 1996).
It is expected that a burst of star formation occurs
at the passage of the ring wave across the disk,
and strong color gradients are indeed observed across the ring.
 In the Cartwheel, the H$\alpha$ comes exclusively from the
outer ring (Higdon 1995), and 80\% of it comes from just
one quadrant, quite asymmetrically. However,
 the quarter of the maximum HI density in the ring does not
coincide with the quarter of maximum star formation (Higdon 1996).
A solution to this problem is to invoke large amounts
of invisible H$_2$ gas, although no CO emission has been detected in
this galaxy (cf Horellou et al 1995). The CO-to-H$_2$ conversion 
ratio could be much higher than standard in this low-metallicity
object. Simulations have been carried out (e.g. Hernquist \& Weil 1993),
but they predict concentrations of gas in the spokes and in the
center, that are not observed.
The collisional ring galaxies are in fact much more complex systems
than they first appear.

Another interesting case is that of polar ring galaxies,
where a large amount of gas has been accreted during
an interaction. The event is old enough that old stars are
observed now in the accreted gas disk. In fact, the accreted
system displays normal abundances for spiral disks
(Eskridge \& Pogge 1997), and appears to be very stable and
long-lived, possibly because of the triaxial structure
(Peletier et al 1993). Dynamical models of polar rings
reveal that the rings are very massive, and even could be the
more massive system (Combes \& Arnaboldi 1996,
Arnaboldi et al 1997). This made Schweizer (1997) to suggest
that the accreted system is the central lenticular (it would
be a "failed" bulge). This is unlikely however, since in a merging
the stellar component is heated and dispersed all around, while 
only the gas can cool down and settle in a perpendicular disk.
There could not have been two gaseous disks in the interaction,
since they would then have settled in a single disk.

\subsection{Mergers}

The merging event can provide one of the most extreme
mixing of matter and elements, while triggering a violent
relaxation. Starbursts are often associated with mergers,
and will contribute to enhance gradients, since a huge
amount of gas is piling up in the center. But in the same
time long gaseous tidal tails are dragged away. The extent
to which gas can be splashed over is spectacular in many
systems: M81/M82 (Yun et al 1993), NGC7252 (Hibbard et al 1994), Arp295,
NGC520, the Mice (Hibbard 1995). Figure 2 is showing how
the fraction of HI gas in the tails increases along a merging
sequence. In fact, most of this gas remains bound to the merged system,
and already some of the bases of the tails are turning back towards the center.
This is done through a 
phase-wrapping process, and shells and ripples are formed.
 Since the HI gas is more and more present in the tails
at more advanced stages, it appears to have little influence
on mixing and abundance gradients. The time-scale to return is a significant 
fraction of the Hubble time. On the contrary, huge molecular
concentrations are observed through the CO lines in mergers,
sometimes up to 50\% of the dynamical mass in the center
(Scoville et al 1994). In general, there is much more CO emission
in interacting galaxies, and it is more concentrated
(e.g. Braine \& Combes 1993). The main result is certainly
to rebuilt abundance gradients in the merging system, which 
will become an early-type or elliptical.

\begin{figure}[t]
% psfile=#1 vsize=#2 angle=#3 hscale=#4 vscale=#5 hoffset=#6 voffset=#7
%\plotfiddle{combes_qc_f2.ps}{65truemm}{0}{50}{50}{0}{0}
\psfig{figure=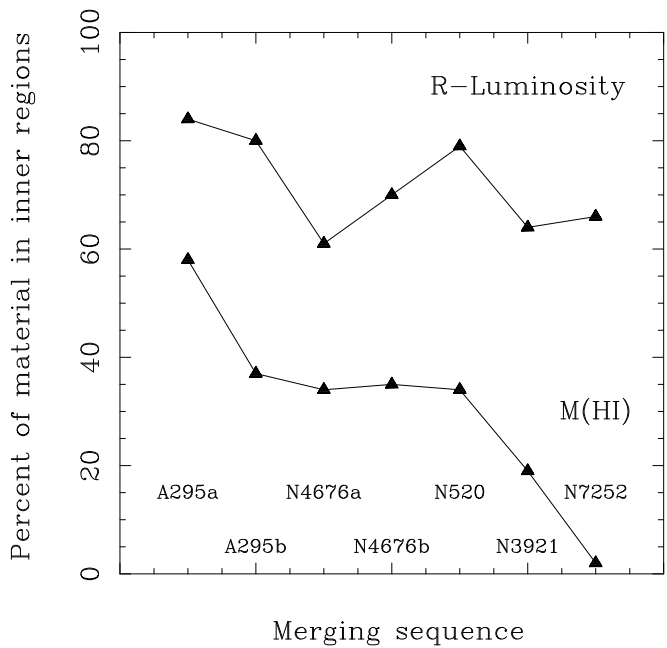,bbllx=1mm,bblly=10mm,bburx=185mm,bbury=190mm,width=9.5cm,angle=0}
\caption{Fraction of light and atomic gas contained within the
inner regions of systems along a merging sequence. Systems at left
are beginning their interaction, while at right they are merged
(NGC 7252). The fraction of the HI gas remaining in the center
drops drastically in the merging process, while the gas is 
dragged out in tidal tails (from Hibbard 1995).}
\end{figure}

Abundances can be determined in the molecular medium
through trace molecules such as CO, CS, CN, HCN, HNC, HCO$^+$
and their isotopes (e.g. Henkel \& Mauersberger 1993).
Nuclei of starburst galaxies reveal anomalous 
isotopic ratios, in particular $^{12}$C and $^{18}$O are 
enhanced. Since high mass stars transform $^{14}$N in $^{18}$O, 
this has been interpreted through an IMF favoring high 
mass stars in starbursts (e.g. Doyon et al 1994).
As a frequent feature of mergers, the 
$^{13}$CO/$^{12}$CO ratio is observed to be 4-5 
times the usual one in disks (Aalto et al 1991).
 This can be explained by infall of fresh gas, which is not 
enriched in the secondary element $^{13}$C 
(Casoli Dupraz \& Combes 1992).

\subsection{Ellipticals: results of mergers?}

Abundance gradients are generally observed also in 
elliptical galaxies (cf Worthey, O'Connell, this meeting). 
The interpretation
of the observations in terms of stellar populations and ages
encounter many difficulties and ambiguities: age-
metallicity degeneracy, dust reddening.
Line indices Mg2, Fe, H$\beta$ can provide more information
through theoretical diagrams (Worthey et al 1994).
These indicate that ellipticals
are old in general (13-15 Gyrs), 
but with very different histories of star 
formation. The duration of the star formation period 
appears inversely proportional to 
the velocity dispersion (Bressan et al 1996).
There might be a dichotomy between spheroids and 
ellipticals (Gorgas et al 1997).

Vazdekis et al (1997), studying 20 absorption lines, 
find good fits for both models: either a
single-age single-metallicity model or a
full chemical evolutionary model. This can be understood
since more than 80\% of the stars form within 1.5 Gyr after the 
formation of the galaxy.

Contrary to what could be expected a priori for
merged systems with a lot of mixing, it is
quite easy to form abundance gradients in 
elliptical galaxies, even with dissipationless collapse 
(Stiavelli \& Matteucci 1991).
The de Vaucouleurs light profile in $r^{1/4}$ is obtained
if ellipticals are formed rapidly through collapse of clumps 
(van Albada 1982, Aguilar \& Merritt 1990). Abundance
gradients build in when considering star formation in the 
clumpy gas to be extended in time, though 
fast (but not instantaneous, as oversimplified).
The presence of dark matter (at least half of 
the total mass), helps the collapse of the gas to 
produce abundance gradients. 
The effect of dark matter (DM) is that galactic winds occur 
much later to prevent star formation.
In the absence of DM, a greater amount of 
dissipation is required to obtain the same 
gradients (Stiavelli \& Matteucci 1991). However,
in dissipative models, it is concluded that DM reduce the 
agreement with observations (e.g. Carlberg 1984).
The gas must be self-gravitating, to obtain 
velocity anisotropy (no DM necessary, 
or less than half the mass).
Models reproduce the observed relation between
metal abundance and mass (or $\sigma$).
The gas physics has an effect only for low mass 
galaxies that are not dominated by gravity.
Under the effects of gas pressure, the system
tends to be more isotropic.

Elliptical systems have in general low rotation,
and radial orbits tend to wash out the gradients.
However, isochromes are smoothed out later 
than isophotes.
More than 20 crossing times are necessary to mix stars born 
at different epochs.
Even if the gas is efficiently smoothed (i.e. all 
stars born at the same time have the same 
metallicity), the abundance gradients can come from spatial 
segregation of the various stellar 
generations (Stiavelli \& Matteucci 1991).
Young and high-Z stars form in more collapsed 
and compact gas in the center, such that 
isochromes are more flattened than isophotes.

%----------------------------------------------------------------------
\section{N-body simulations}

Simulations help to understand the mixing effects of galaxy
interactions. In fact, they show that the violent relaxation
is never complete, so that 
the pre-existing abundance gradients in 
the merging spirals remain (Mihos \& Hernquist 1994b).
Moreover, abundance gradients are enhanced through violent 
starbursts.

\subsection{Modelisation of the star formation}

The star formation rate and IMF are not well known, especially 
in perturbed circumstances, therefore this  
modelisation is at best exploratory.
Taking into account the star formation is however 
essential for the dynamics: there is
energy released in the gas, some
gas mass locked up into the dissipationless component,
which slows down the inward mass flow.
In the literature, the star formation rate has been 
derived from cloud-cloud collisions (Noguchi \& Ishibashi 1986),
or more generally from the empirical
Schmidt law (Mihos et al  1992), with an
exponent $n \sim$ 1.5- 2, i.e. a
non-linear law with density.
In every cases, tidal interaction and mergers 
increase the SFR transiently by an order of 
magnitude, because of the concentration of the gas mass in 
small regions. The efficiency
strongly depends on the exponent $n$ of the Schmidt law.
Simulations have not enough dynamical range to deal
properly with star formation (a single particle is of
the order of 10$^6$ M$_\odot$), and simple recipes
have been adopted to transform gas in stars and 
reciprocally (Mihos \& Hernquist 1994a): the same 
particles are X\% stars and (1-X)\% gas for instance.
This couples young stars and gas kinematics
artificially, and  also inhibits contagious star formation,
or feedback mechanisms (Struck-Marcell \& Scalo 1987).

\subsection{Gas flow and starburst triggering}

In major merger simulations, strong non-
axisymmetric forces are exerted on the gas.
The main torques responsible for the gas flow 
are not directly due to the companion,
but to the tidally- triggered perturbations 
on the primary disk: bars, spirals.
Self-gravity plays the essential role:
internal perturbations are generated on the primary 
gas disk (the gas accreted from the companion is
not predominant).
The essential parameter is the bulge-to-disk ratio in the 
primary, regulating the instabilities
(Mihos \& Hernquist 1996).
With large-bulge galaxies, instabilities
are delayed, and the interaction ends with
a strong starburst. In small-bulge galaxies, 
there is continuous activity, weaker at the 
end when the gas has been consumed up.
The geometry of the encounter has only minor effects
(external manifestations such as tails, debris).
Globally about 75\% of the gas is 
consumed during the merger.
The tidal tails gas will rain down, in 1 Gyr 
scale, but in the outer parts essentially.

%----------------------------------------------------------------------
\section{Conclusion}
Paradoxically, the mixing is not predominant
in interactions and mergers. On the contrary abundance
gradients are enhanced in such events,
because of gas concentration and starburst,
except for very rare cases of mergers of pure 
stellar systems.

We have shown that the angular momentum is tranfered
essentially through gravity torques, and therefore
the abundances are strongly coupled to the dynamics.
Star formation triggering and efficiency are not
yet well known in interacting galaxies. It turns
out quite easy to build abundance gradients in 
ellipticals, either through rapid initial formation,
merger, or less violent evolution, so the observed
gradients are not a stringent test of the 
formation processes. 

%----------------------------------------------------------------------
\acknowledgments 
My warmful thanks go to the organisers for this very interesting
and interactive meeting.
%----------------------------------------------------------------------

%----------------------------------------------------------------------
\end{document}